\documentclass[11pt,fleqn]{article}

\usepackage{amsmath,amssymb,amsthm,cite,graphics,epsfig,enumerate}

\newcommand{\thetbn}{\arabic{nomer}}

\newcommand{\todo}[1][\null]{\ensuremath{\clubsuit}}
\newcommand{\checked}[1][\null]{\ensuremath{\diamond}}
\newcommand{\noprint}[1]{}

\newcounter{tbn}

\newcounter{mcasenum}

\newtheorem{theorem}{Theorem}

\newtheorem{corollary}{Corollary}

\newtheorem*{proposition*}{Proposition}
{\theoremstyle{definition}

\newtheorem{remark}{Remark}

}

\begin{document}\allowdisplaybreaks
\begin{center}
\par\noindent {\LARGE\bf
Equivalence groupoid of a class of variable coefficient Korteweg--de Vries  equations
\par}

{\vspace{5mm}\par\noindent\large  Olena~Vaneeva$^{\dag }$ and Severin Po{\v s}ta$^{\ddag}$
\par\vspace{1mm}\par}
\end{center}

{\par\noindent\it\small
${}^\dag$\ Institute of Mathematics of the National Academy of Sciences of Ukraine,\\[1ex]
$\phantom{{}^\ddag}$\ 3 Tereshchenkivska Str., 01601 Kyiv-4, Ukraine\\[1ex]
${}^\ddag$\ Department of Mathematics, Faculty of Nuclear Sciences and Physical Engineering,\\[1ex]
$\phantom{{}^\ddag}$\ Czech Technical University in Prague, 13 Trojanova Str., 120 00 Prague, Czech Republic
}

{\vspace{2mm}\par\noindent
$\phantom{{}^\dag{}\;}\ $E-mails: \it vaneeva@imath.kiev.ua,
severin.posta@fjfi.cvut.cz
\par}

{\vspace{5mm}\par\noindent\hspace*{5mm}\parbox{150mm}{\small
We classify the admissible transformations in a class of variable coefficient Korteweg--de Vries  equations. As a result, full description of the structure of
the equivalence groupoid of the class is given. The class under study is partitioned into six disjoint normalized subclasses. The widest possible equivalence group  for each subclass is found which appears to be generalized extended in five cases. Ways for improvement of transformational properties of the subclasses are proposed using gaugings of arbitrary elements and mapping between classes. The group classification of one of the subclasses is carried out as an illustrative example.
}\vspace{4mm}\par}

\section{Introduction}

 It is widely known that there is no general theory for integration of  nonlinear partial differential equations (PDEs).
Nevertheless, many special cases of complete integration or finding particular solutions are related to appropriate changes of variables.
Nondegenerate point transformations that leave a differential equation invariant and form a
connected Lie group  are called Lie symmetries of this equation. Transformations of this kind are ones which are mostly
used. In many cases,
 the algorithmic Lie reduction method, which uses known Lie symmetries, results in the construction of  group-invariant
solutions for a given PDE. 
This places
the transformation methods  among the most powerful analytic tools currently
available in the study of nonlinear PDEs~\cite{soph_thesis}.

Many  nonlinear PDEs that are important for applications are parameterized by arbitrary elements (constants or functions) and constitute classes of PDEs.
An important task is to study transformational properties of such classes, i.e. to describe explicitly nondegenerate point transformations that link members of the class.  Indeed,
if two differential equations are connected by such a  transformation,
then associated objects like exact solutions, local conservation laws, and different kinds of symmetries of these equations are also related by this transformation. Such equations are called equivalent (or {\it similar} in terms of~\cite{ovsiannikov}).
Knowledge of an exact solution for one of two equivalent equations allows one to construct the corresponding exact solution for the other  equation  using the  point transformation connecting them. A number of exact solutions for variable coefficient PDEs were constructed using the equivalence method, see, e.g.,~\cite{vane2009,kuri2015,popo2010b}.
At the same time,
nondegenerate point transformations appear to be a useful tool
not only for
finding exact solutions but also for exhaustive solving group classifications problems~(see, e.g.,~\cite{popo2010a,kuri2015,bihlo2012b} and reference therein), design of physical parameterization schemes~\cite{popo2012a}, and study of integrability~\cite{vane2014phys,joshi,gurses,kundu,brugarino}.

The systematic study of transformational properties of classes of nonlinear PDEs was initiated in 1991 by Kingston and Sophocleous~\cite{king1991a,kingston1991,soph_thesis}.
These authors later named the transformations related two particular equations in a class of PDEs  {\it form-preserving transformations}~\cite{kingston1997}, because such transformations preserve the form of the equation in a class and change only its arbitrary elements.
Only a year later in 1992 Gazeau and Winternitz started to investigate such transformations in classes of PDEs calling them {\it allowed transformations}~\cite{Gazeau1992,wint1992}. Rigorous definitions and developed theory on the subject was proposed later by Popovych~\cite{popo2010a,popo2006}. As formalization of notion of form-preserving (allowed) transformations the term {\it admissible transformation} was suggested therein. In brief, an admissible transformation is   a triple consisting of two fixed equations from a class
and a transformation that links these equations.
The set of admissible  transformations considered with the standard operation of composition of  transformations is also called the equivalence groupoid. The necessary notions and concepts are presented in more detail in the next section.

Some recent studies on admissible transformations (equivalence groupoid) of physically interesting classes of PDEs, such as variable coefficient Burgers, Kawahara and Schr\"odinger equations, can be found in particular in~\cite{poch2015,kuru2015,poch2014,kuri2015}.

Notably that one of the first classes whose transformational properties were investigated was the class of
remarkable variable-coefficient Korteweg--de Vries equations
\begin{equation}\label{vc_KdVtx}
\mathcal L:\quad u_t+f(t,x)uu_x+g(t,x)u_{xxx}=0,\quad fg\neq0.
\end{equation}
The classical Korteweg--de Vries   equation and its generalizations model various  physi\-cal systems, including
classical water waves, gravity waves, plasma waves and waves in lattices~\cite{jeffrey,jeffrey_book,grimshaw,cascaval}.
There are a number of works, where equations from the class~$\mathcal L$ were investigated from various points of view.
We mention only the works most related to our study. In papers~\cite{joshi,hlavaty} the Painlev\'e analysis  of equations~\eqref{vc_KdVtx}
was performed. It was shown that such equations pass  Painlev\'e test only if the coefficients $f$ and $g$ satisfy the conditions  $f_x=g_x=0,$ and $(g/f)_t/f={\rm const.}$ These conditions coincide with those of reducibility of variable coefficient equations~\eqref{vc_KdVtx} to the standard Korteweg--de Vries   equation, which is foreseeable.

Lie symmetries and allowed transformations of equations from the class~$\mathcal L$ were studied in~\cite{Gazeau1992,wint1992,gungor1996}.
We aim to present enhanced results on  transformational properties of these equations, which will be investigated with  modern point of view. It was deduced correctly in~\cite{Gazeau1992,wint1992} that there are six different subclasses of the class~\eqref{vc_KdVtx} which differ by their transformational properties.
Nevertheless, there are several reasons to reconsider the problem of classification of admissible transformations in this class:
\begin{itemize}
\item The consideration in~\cite{Gazeau1992,wint1992} was restricted to fiber-preserving point transformations without a proof that transformations of such kind exhaust all possible admissible point transformations.
We aim to show that not only point but even contact admissible transformations are exhausted by fiber-preserving ones.

\item When six subclasses that differ by their transformational properties were derived  in~\cite{Gazeau1992,wint1992} the arbitrary elements were gauged
whenever possible from the very beginning.
The admissible transformations were indicated
for the simplified (gauged) forms of equations. Our strategy is firstly to consider the subclasses without simplification and only then  the gauged ones.

\item Only transformation components for independent and dependent variables were adduced in~\cite{Gazeau1992,wint1992} without rules for changing arbitrary elements.
We formulate results in terms of  equivalence groups of  appropriate kinds (usual or generalized extended), where transformations for both independent and dependent variables $t,$ $x$, $u$ and arbitrary
elements $f$, $g$ are presented.

\item Some admissible transformations were found in~\cite{Gazeau1992,wint1992} in an implicit form, namely they included functions that satisfy certain ordinary differential equations. We aim to present all the results in an explicit form.
\end{itemize}

The concept of normalized classes allows us to explain why subclasses with different transformational properties arise in the course of the investigation. In fact they are maximal normalized subclasses in the class under study. Therefore, all the consideration within the modern approach becomes complete and clear.

The structure of the paper is as follows. Firstly we recall the necessary definitions and  notions on admissible transformations in classes of differential equations. In section 3 complete classification of admissible transformations in the class~\eqref{vc_KdVtx} is performed. The structure of its equivalence groupoid is described exhaustively in terms of partitioning into six disjoint normalized subclasses.  In section 4 we present the results on transformational properties of the gauged subclasses of the class~\eqref{vc_KdVtx}. In section~5 we show how the method
of mapping between classes proposed in~\cite{vane2009} can be used to improve transformational properties of one of the derived subclasses. Using this method the group classification problem for this subclass is solved as an illustrative example. The results on admissible transformations are summarized  in Conclusion.

\section{Basic definitions}

In this section we give basic definitions on admissible transformations in classes of differential equations, following the papers~\cite{popo2006,popo2010a} in general outlines.

Let a system of differential equations consists of $l$ equations with independent variables $\mathrm x=(x_1,\dots, x_n)$ and dependent variables $u=(u^1,\dots,u^m)$, which include also $k$ arbitrary elements  $\sigma(\mathrm x,u_{(p)}) = (\sigma^1(\mathrm x,u_{(p)}),\dots,\sigma^k(\mathrm x,u_{(p)}))$. Here the notation~$u_{(p)}$ includes the tuple of dependent variables~$u=(u^1,\dots,u^m)$ and all the derivatives of~$u$ with respect to~$\mathrm x$ of order up to~$p$.
We denote such a system as $\mathcal L_\sigma$: \smash{$L(\mathrm x,u_{(p)},\sigma_{(q)}(\mathrm x,u_{(p)}))=0$}.
Let~$\sigma_{(q)}$  denote the partial derivatives of the arbitrary elements $\sigma$ of order not exceeding $q$ for which both $\mathrm x$ and $u_{(p)}$ act as the independent variables. The tuple of arbitrary elements~$\sigma$ runs through the solution set of the system of auxiliary differential equations $S(\mathrm x,u_{(p)},\sigma_{(q')}(\mathrm x,u_{(p)}))=0$ and inequalities $\Sigma(\mathrm x,u_{(p)},\sigma_{(q')}(\mathrm x,u_{(p)}))\ne0$,
in which again both~$\mathrm x$ and~$u_{(p)}$ act as independent variables. Components of the tuples $S$ and~$\Sigma$ are smooth functions of~$\mathrm x$, $u_{(p)}$ and $\sigma_{(q')}$.

Then the set $\{\mathcal L_\sigma\mid\sigma\in\mathcal S\}$ denoted by~$\mathcal L|_{\mathcal S}$ is called a \emph{class of differential equations}
defined by parameterized systems~$\mathcal L_\sigma$ and the set~$\mathcal S$ of arbitrary elements~$\sigma$.

For the class $\mathcal L$ of variable coefficient Korteweg--de Vries equations~\eqref{vc_KdVtx} there are two independent variables $\mathrm x=(t,x)$, $n=2$, one dependent variable $u$, $m=1$, and two arbitrary elements $f$ and $g$, therefore $k=2$ and $\sigma=(f,g)$.
The auxiliary system  $S=0$ and $\Sigma\ne0$  consists of the equations $f_u=f_{u_t}=f_{u_x}=\dots =f_{u_{xxx}}=0$, the same system of equations for the function $g$, and the inequalities $f\neq0$ and $g\neq0.$

Now we recall
the notion  of admissible transformations and the equivalence groupoid adopting them to our case.
Given equations $\mathcal L_\sigma$ and \smash{$\mathcal L_{\tilde\sigma}$} singled out from the class~$\mathcal L|_{\mathcal S}$ by fixing two values of  the arbitrary elements $\sigma,\tilde\sigma\in\mathcal S$, the set $ \mathcal T(\sigma,\tilde \sigma)$ of point transformations that map the equation $\mathcal L_\sigma$ to the equation \smash{$\mathcal L_{\tilde\sigma}$} is called the set of admissible point transformations from $\mathcal L_\sigma$ to \smash{$\mathcal L_{\tilde\sigma}$}.
In particular, the maximal point symmetry (pseudo)group $G_\sigma$ of the system $\mathcal L_\sigma$ coincides, by definition, with the set $\mathcal{T}(\sigma,\sigma)$ of point transformations that leave the equation $\mathcal L_\sigma$ invariant.

An ordered triplet $(\sigma,\tilde\sigma,\phi)$, where $\sigma,\tilde\sigma\in\mathcal S$ with $\mathcal T(\sigma,\tilde\sigma)\ne\varnothing$ and $\phi\in \mathcal T(\sigma,\tilde\sigma)$, is called an \emph{admissible transformation} in the class~$\mathcal L|_{\mathcal S}$.
Speaking the simpler way, an admissible transformation is a triplet consisting of
a source equation, a target equation, and a point transformation connecting these two equations.

The set of admissible transformations of the class~$\mathcal L|_{\mathcal S}$ considered with the operations of composition and taking the inverse
has a structure of groupoid and therefore is called the \emph{equivalence groupoid} $\mathcal G^\sim=\mathcal G^\sim(\mathcal L|_{\mathcal S})$  of the class~\cite{popo2012a}.
Indeed, the partial binary operation of composition is naturally defined for pairs of admissible transformations
for which the target equation of the first admissible transformation and the source equation of the second admissible transformation coincides,
$(\sigma,\tilde\sigma,\phi)\circ(\tilde\sigma,\bar\sigma,\tilde\phi)=(\sigma,\bar\sigma,\tilde\phi\circ\phi)$.
Each admissible transformation is invertible, $(\sigma,\tilde\sigma,\phi)^{-1}=(\tilde\sigma,\sigma,\phi^{-1})$,
i.e., the inversion of admissible transformations is a unitary operation defined everywhere.
Therefore, all the groupoid axioms are obviously satisfied for~$\mathcal G^\sim$~\cite{bihlo2015}.

For a general class of differential equations,
it is not possible to relate its equivalence groupoid with a group of point transformations
that act in the extended space of $(\mathrm x,u_{(p)},\sigma)$, respect the contact structure on the space of $(\mathrm x,u_{(p)})$ and preserve the class.
But it is possible to single out a maximal part of the equivalence groupoid that meets such requirements,
then the associated group is called the equivalence group $G^{\sim}$ of the class.

Elements of $G^{\sim}$ are called \emph{equivalence transformations} of the class~$\mathcal L|_{\mathcal S}$.
Each equivalence transformation~$\Phi$ induces a family of admissible transformations parameterized by the arbitrary elements,
$\{(\sigma,\Phi\sigma,\Phi|_{(\mathrm x,u)})\mid \sigma\in\mathcal S\}$.

There exist several kinds of equivalence groups depending on restrictions that are imposed on the transformations.
The \emph{usual equivalence group} of the class~$\mathcal L|_{\mathcal S}$
consists of the nondegenerate point transformations in the space of variables and arbitrary elements,
which preserve the whole class~$\mathcal L|_{\mathcal S}$ and are projectable on the variable space,
i.e., the transformation components corresponding to independent and dependent variables do not depend on arbitrary elements~\cite{ovsiannikov}.

For the class~\eqref{vc_KdVtx} the projectibilty condition means that the transformation components for $t,$ $x,$ and $u$ of usual equivalence transformations do not depend on arbitrary elements of the class and have the form
$
(\tilde t,\tilde x,\tilde u)=(T,X,U)(t,x,u).
$

In some cases, e.g., if the arbitrary elements~$\sigma$ do not depend on derivatives of $u$,
we can neglect the condition that the transformation components for~$(t,x,u)$ of equivalence transformations do not involve~$\sigma$,
which gives the \emph{generalized equivalence group} $G^{\sim}_{\rm gen}=G^{\sim}_{\rm gen}(\mathcal L|_{\mathcal S})$
of the class~$\mathcal L|_{\mathcal S}$ \cite{popo2006,popo2010a,mele94Ay}.
Each element~$\Phi$ of~$G^{\sim}_{\rm gen}$ is a point transformation in the $(\mathrm x,u,\sigma)$-space
such that for any~$\sigma$ from~$\mathcal S$ its image $\Phi\sigma$ also belongs to~$\mathcal S$,
and $\Phi(\cdot,\cdot,\sigma(\cdot,\cdot))|_{(\mathrm x,u)}\in\mathcal{T}(\sigma,\Phi\sigma)$.

Therefore, for the class~\eqref{vc_KdVtx} the  transformation components for $t,$ $x,$ and $u$  of generalized equivalence transformations are of the form
$
(\tilde t,\tilde x,\tilde u)=(T,X,U)(t,x,u,f,g).
$
The generalized transformations are nontrivial if and only if at least one of the following inequalities holds,
\[
\left(\dfrac{\partial T}{\partial f},\dfrac{\partial X}{\partial f},\dfrac{\partial U}{\partial f}\right)\neq(0,0,0)\quad\mbox{and}\quad\left(\dfrac{\partial T}{\partial g},\dfrac{\partial X}{\partial g},\dfrac{\partial U}{\partial g}\right)\neq(0,0,0).
\]

If transformations from the equivalence group appear to depend on arbitrary elements of the class in a nonlocal way, then such an equivalence group is called {\it extended} one~\cite{ivan2005,popo2010a}. Nevertheless, the transformations should at least be point with respect to the independent and the dependent variables
while the arbitrary elements are fixed.

In the case when both  the conditions of projectability and locality of equivalence transformations are simultaneously weaken,
 the notion of \emph{generalized extended equivalence group}   naturally appears.

The class~$\mathcal L|_{\mathcal S}$ is called normalized in the usual  (resp. generalized, resp. extended, resp. generalized extended)  sense
if the equivalence groupoid of this class is induced by  transformations from its equivalence group of the corresponding type~\cite{popo2006,popo2010a}. The normalization property becomes crucial in particular for solving group classification problems via the algebraic method~\cite{bihlo2012b,bihlo2015,boyko2015}.

\section{Classification of admissible transformations}

Though the study of transformational properties in classes of PDEs are more often restricted to point admissible transformations it is possible also to consider contact admissible transformations and the respective contact equivalence groupoid.
It was shown in~\cite{vane2014phys} that for the normalized class of evolution equations of the form  $\bar{\mathcal L}\colon \ u_t=F(t,x,u,u_x)u_n+G(t,x,u,u_1,\dots,u_{n-1})$ for $n\geq3$ its contact equivalence groupoid coincides with the respective point one, i.e., any contact admissible transformation between two equations from the class $\bar{\mathcal L}$
is the first prolongation of a~point transformation between these equations.
Therefore, for subclass~\eqref{vc_KdVtx} of the class $\bar{\mathcal L}$ it suffices to investigate only point admissible transformations.

It is well known
that any point transformation~$\mathcal T$ relating two fixed evolution equations in 1+1 dimensions  has the form
$\tilde t=T(t)$, $\tilde x=X(t,x,u)$, $\tilde u=U(t,x,u)$ with $T_t(X_xU_u-X_uU_x)\neq0$. The partial derivatives involved in equations~\eqref{vc_KdVtx} are transformed as follows:
\[\tilde u_{\tilde t}=\frac1{T_t}\left({D_tU}-\frac{D_tX}{D_xX}D_xU\right),\quad \tilde u_{\tilde x}=\frac{D_xU}{D_xX},\quad \tilde u_{\tilde x\tilde x\tilde x}=\frac1{D_xX}D_x\left(\frac1{D_xX}D_x\left(\frac{D_xU}{D_xX}\right)\right),\]
where $D_t=\partial_t+u_t\partial_{u}+u_{tt}\partial_{u_t}+u_{tx}\partial_{u_x}+\dots{}$ and
$D_x=\partial_x+u_x\partial_{u}+u_{tx}\partial_{u_t}+u_{xx}\partial_{u_x}+\dots{}$
are operators of the total differentiation with respect to~$t$ and~$x$. Here and below subscripts of functions indicate partial derivatives with respect to the corresponding variables.

We substitute these expressions into the equation  $\tilde u_{\tilde t}+\tilde f(\tilde t,\tilde x)\tilde u\tilde u_{\tilde x}+\tilde g(\tilde t,\tilde x)\tilde u_{\tilde x\tilde x\tilde x}=0$ and obtain an equation in terms of the untilded variables. In order to confine this equation to the manifold defined by~\eqref{vc_KdVtx} in the third-order jet space
with the independent variables $(t,x)$ and the dependent variable~$u$
we further substitute into it the expression $u_t=-f(t,x)u u_x-g(t,x)u_{xxx}$.
Splitting the obtained identity with respect to the derivatives of $u$ leads to the determining
equations on the functions~$T$, $X$, and $U$. In particular, the coefficient of $u_{xxx}$ is $(\tilde g T_t-g(X_x+X_uu_x)^3)(X_xU_u-X_uU_x)$.
Setting it to zero and taking into account the nondegeneracy condition $T_t(X_xU_u-X_uU_x)\neq0$ we get $\tilde g T_t-g(X_x+X_uu_x)^3=0$. Further splitting with respect to $u_x$ leads to the conditions $X_u=0$ and $\tilde g T_t=g X_x^3.$ Therefore, $X=X(t,x)$ and it means that there are no other admissible transformations except fiber-preserving ones. This agrees with the results of papers~\cite{vane2014phys,kingston1997}, derived for more general classes of evolution equations.

The condition $X_u=0$ leads to essential simplification of the determining equations, thus we get additional constraints for the coefficients $X$ and $U$:  $U_{uu}=U_{xu}=X_{xx}=0.$  Therefore, the transformation components for independent and dependent variables of admissible transformations have the form
\begin{equation}\label{eqXU}
\tilde t=\theta(t), \quad \tilde x=\alpha(t)x+\beta(t),\quad \tilde u=\varphi(t)u+\psi(t,x),
\end{equation}
where $\theta$, $\alpha$, $\beta$, $\varphi$ and $\psi$ are arbitrary smooth functions of their variables with $\dot\theta\alpha\varphi\neq0.$

The formulas \begin{gather}\label{eq_adm_tr}
\begin{array}{l}
\tilde f=\dfrac{\alpha}{\dot\theta\varphi} f, \quad
\tilde g=\dfrac{\alpha^3}{\dot\theta} g
\end{array}
\end{gather}
establish the connection between values of arbitrary elements of the initial equation and the target equation. We denote by overdots ordinary derivatives with respect to the variable $t$.

The classifying equations involving $\theta,$ $\alpha,$ $\beta$, $\varphi$ and $\psi$ and the arbitrary functions $f$ and $g$ are of the form
\begin{gather}\label{clas_eq_adm_tr1}
f\psi_x+\dot\varphi=0, \\[1ex]\label{clas_eq_adm_tr2} f\alpha \psi =\varphi(\dot\alpha x+\dot\beta),\\[1ex]\label{clas_eq_adm_tr3}
\psi_t+g\psi_{xxx}=0.
\end{gather}

To deduce the equivalence group of the class~\eqref{vc_KdVtx} equations~\eqref{clas_eq_adm_tr1}--\eqref{clas_eq_adm_tr3} should be split with respect to arbitrary elements $f$ and $g$. This results in the equations
\begin{gather*}
 \dot\alpha=\dot\beta=\dot\varphi =\psi=0.
\end{gather*}
Therefore, the following statement is true.
\begin{theorem}
The class $\mathcal L$ of equations~\eqref{vc_KdVtx}
admits the usual equivalence group~$G^{\sim}$ consisting of the transformations:
\begin{gather*}
\tilde t=\theta(t),\quad \tilde x=\delta_1x+\delta_2,\quad\tilde u=\delta_3u,  \quad
\tilde f(\tilde t,\tilde x)=\dfrac{\delta_1}{\delta_3\dot\theta(t)} f(t,x), \quad
\tilde g(\tilde t,\tilde x)=\dfrac{\delta_1{}^3}{\dot\theta(t)} g(t,x),
\end{gather*}
where~$\theta$ is an
 arbitrary function of~$t$ with $\dot\theta\neq0$,
$\delta_j$ $(j=1,2,3)$ are arbitrary constants with $\delta_1\delta_3 \not=0$.
\end{theorem}

The class~$\mathcal L$ is not normalized in any sense, there exist subclasses of this class singled out by setting restrictions on values of arbitrary elements $f$ and $g$ which possess wider equivalence groups than the group~$G^{\sim}$.  These subclasses and the corresponding {\it maximal conditional equivalence groups}~\cite[Definition 7]{popo2010a} can be found investigating the system~\eqref{clas_eq_adm_tr1}--\eqref{clas_eq_adm_tr3}.

Equation~\eqref{clas_eq_adm_tr1} results in $\psi=\dfrac1f\dfrac{\varphi}{\alpha}\left(\dot\alpha x+\dot\beta \right)$, then    equation~\eqref{clas_eq_adm_tr2} takes the form
\begin{gather*}
   \left(\frac{\dot\alpha }{\alpha}x+\frac{\dot\beta }\alpha\right){f_x}=\left(\frac{\dot\varphi}\varphi+\frac{\dot\alpha}{\alpha}\right)f.
\end{gather*}
There are only three possibilities for the nonvanishing function $f(t,x)$ to satisfy this equation:
I. $f=f(t)\neq0$; II. $f=p(t)e^{q(t)x}$, $pq\neq0$, and III. $f=p(t)(x+q(t))^{r(t)},$ $pr\neq0.$ Further we consider each case separately.

\medskip

{\bf I.} $f=f(t)\neq0$. We substitute this form of $f$ into the equations~\eqref{clas_eq_adm_tr1}--\eqref{clas_eq_adm_tr3}. This results in the conditions
$\psi_t=\psi_{xx}=0$ and therefore $\psi= c_1x+c_2$, where $c_1$ and $c_2$ are arbitrary constants. Equation~\eqref{clas_eq_adm_tr3} becomes an identity, hence
there is no restriction on the form of the function $g$ and it remains arbitrary. Equations~\eqref{clas_eq_adm_tr1},~\eqref{clas_eq_adm_tr2} imply the system of equations for $\alpha$, $\beta$ an $\varphi$ of the form: $\dot\varphi =-c_1f,$ $\varphi\dot\alpha =c_1f\alpha$,
and $\varphi\dot\beta=c_2f\alpha.$ The general solution of this system and formulas~\eqref{eq_adm_tr} give the full description of admissible transformations
of the subclass of class~\eqref{vc_KdVtx} with $f_x=0$. After redenotion of constants $c_1=-\delta_3$, $c_2=-\delta_2\delta_3/\delta_1$ we get the following assertion.

\begin{theorem}The generalized extended equivalence group~$\hat G^{\sim}_1$ of the class
\begin{equation}\label{vc_KdVtx_fx=0}
\mathcal L_1:\quad u_t+f(t)uu_x+g(t,x)u_{xxx}=0,
\end{equation}
 consists of the transformations
  \begin{gather*}
  \tilde t=\theta(t),\quad \tilde x=\frac{\delta_1x+\delta_2}{\delta_3\int\! f(t)\,{\rm d}t+\delta_4}+\delta_5,\quad \tilde u=\left(\delta_3\int\!\! f(t)\,{\rm d}t+\delta_4\right)u-\delta_3x-\frac{\delta_2\delta_3}{\delta_1},\\
\tilde f(\tilde t)=\frac{\delta_1}{\dot\theta(t)(\delta_3\int\! f(t)\,{\rm d}t+\delta_4)^2}f(t),\quad \tilde g(\tilde t,\tilde x)=\frac{\delta_1{}^3}{\dot\theta(t)(\delta_3\int\! f(t)\,{\rm d}t+\delta_4)^3}\,g(t,x),
\end{gather*}
where~$\theta$ is an
 arbitrary function of~$t$ with $\dot\theta\neq0$, and $\delta_j$ $(j=1,2,3,4,5)$ are arbitrary constants with $\delta_1(\delta_3{}^2+\delta_4{}^2)\neq0$.

\end{theorem}
The usual equivalence group of the class~\eqref{vc_KdVtx_fx=0} coincides with the equivalence group $G^\sim$ of its superclass~\eqref{vc_KdVtx}.

\medskip
{\bf II.} $f=p(t)e^{q(t)x}$, where $pq\neq0.$ In this case we have in fact reparametrization of the class, thus instead of arbitrary element $f(t,x)$ there are two arbitrary elements $p(t)$ and $q(t)$. The equivalence relations for them are given by $\tilde q=\dfrac q\alpha,$ and $\tilde p =\dfrac{p\alpha}{\varphi \dot\theta}\exp(-q \beta /\alpha)$. Equations~\eqref{clas_eq_adm_tr1}--\eqref{clas_eq_adm_tr3} lead to the conditions $\psi=\dfrac{\dot\varphi}{pq}e^{-q x}$, $\dot\alpha=0,$ $\dot\beta=
\dfrac{\dot\varphi\alpha}{\varphi q}$. The function $g$ has the form $g=\dfrac{s(t)- {\dot q}x}{q^3}$ with the additional constraint $s=\dfrac{\ddot\varphi}{\dot\varphi}-\dfrac{\dot p}{p}-\dfrac{\dot q}{q}.$ The function $g$ is also reparameterized now and the function $s$ can be regarded as  another arbitrary element. The admissible transformations
in the deduced subclass are described by the following statement.
\begin{theorem}
The class of equations
\begin{equation}\label{vc_KdVtx_fx_exp}
\mathcal L_2:\quad u_t+p(t)e^{\,q(t)x}uu_x+\dfrac{s(t)- {\dot q(t)}x}{q(t)^3}u_{xxx}=0
\end{equation}
admits the generalized extended equivalence group~$\hat G^{\sim}_2$ consisting of the transformations
\begin{gather*}
\tilde t=\theta(t),\quad \tilde x=\delta_1x+\beta(t),\quad\tilde u=\varphi(t) u+\frac {\dot\varphi(t)}{p(t)q(t)}e^{-q(t)x},  \\
\tilde p(\tilde t)=\dfrac{\delta_1}{\dot\theta(t)\varphi(t)}e^{-\frac{1}{\delta_1}\beta(t)q(t)}p(t),\quad \tilde q(\tilde t)=\frac {q(t)}{\delta_1},\quad \tilde s(\tilde t)=\frac1{{\delta_1}\dot\theta(t)}\left(\delta_1s(t)+{\dot q(t)}\beta(t)\right),
\end{gather*}
where~$\theta$ is an
 arbitrary function of~$t$ with $\dot\theta\not=0$,
 \[\beta(t)=\delta_1\delta_3\int \frac{p(t)e^{\int\! s(t)\,{\rm d}t}}{\varphi(t)}\, {\rm d}t+\delta_2,\quad \varphi(t)=\delta_3\int\! p(t)q(t)e^{\int\! s(t)\,{\rm d}t} {\rm d}t+\delta_4.\]
Here $\delta_j$ $(j=1,2,3,4)$ are arbitrary constants, $\delta_1(\delta_3{}^2+\delta_4{}^2)\not=0$.

\end{theorem}

\medskip
{\bf III.} $f=p(t)(x+q(t))^{r(t)}$ with $pr\neq0.$ In this case equation~\eqref{clas_eq_adm_tr2} implies   $\psi= \dfrac{\varphi(\dot\alpha x+\dot\beta)}{\alpha p(x+q)^r}$, then~\eqref{clas_eq_adm_tr1} leads to the conditions $\dfrac{\dot\varphi}{\varphi}=(r-1)\dfrac{\dot\alpha}{\alpha}$ and $\dot\beta= q\dot\alpha$. In view of the latter condition the function $\psi$ can be rewritten as $\psi=\dfrac{\varphi}p\dfrac{\dot\alpha}\alpha(x+q)^{1-r}$. Then the equation~\eqref{clas_eq_adm_tr3} containing $g$ takes the form $r(r^2-1)g=(x+q)^{3}\left(s-\dot r \ln(x+q)+\dfrac{(1-r)\dot q}{x+q}\right)$, where the function~$s$, that can be regarded as new arbitrary element, satisfies the equation $s\dfrac{\varphi}p\dfrac{\dot\alpha}\alpha=\dfrac{\rm d}{{\rm d}t}\left(\dfrac{\varphi}p\dfrac{\dot\alpha}\alpha\right)$.
It is easy to see that in two special cases, namely, when $r=1$ or $r=-1$, the function $g$ remains arbitrary.
We consider the cases $r\neq\pm1,$ $r=1$ and $r=-1$ separately. The general  solution of the determining equations gives in each of these three cases full description of admissible transformations in the corresponding subclass of the class~\eqref{vc_KdVtx}. The results are collected in Theorems 4--6. We note that in the case $r=-1$ the nontrivial equivalence group which is wider than $G^\sim$ exists only if the functions  $q(t)=c$, where $c$ is an arbitrary constant. In this case we use one more reparameterization introducing the new function $k=1/p,$ so $f$ takes the form $f=k(t)/(x+c).$

\begin{theorem}
The generalized extended equivalence group~$\hat G^{\sim}_3$ of the class
\begin{gather*}
\mathcal L_3\colon\quad u_t+p(t)(x+q(t))^{r(t)}uu_x+{}\\\qquad\qquad \dfrac{(x+q(t))^{3}}{r(t)(r(t)^2-1)}\left(s(t)-\dot r(t)\ln(x+q(t))+\dfrac{(1-r(t))\dot q(t)}{x+q(t)}\right)u_{xxx}=0,\quad r\neq0,\pm1,
\end{gather*}
consists of the transformations
\begin{gather*}
\tilde t=\theta(t),\quad \tilde x=\alpha(t) x+\beta(t),\quad\tilde u=\varphi(t) u+\delta_1e^{\int\! s(t)\, {\rm d}t} (x+q(t))^{1-r(t)},  \\
\tilde p(\tilde t)= \frac{\alpha(t)^{1-r(t)}}{\phi(t)\dot\theta(t)}{p(t)} ,\quad \tilde q(\tilde t)= q(t)\alpha(t)-  {\beta(t)},\\
\tilde s(\tilde t)=\frac1{\dot\theta(t)}\!\left(s(t)+\dot r(t)\ln\alpha(t)\right),\quad\tilde r(\tilde t)=r(t),
\end{gather*}
where~$\theta$ is an
 arbitrary nonvanishing function of~$t$, $\dot\theta \not=0$,
 \begin{gather*}
 \varphi(t)=\delta_1\int\! p(t)(r(t)-1)\,e^{\int\! s(t)\,{\rm d}t}{\rm d}t+\delta_2,\\
 \alpha(t)=\delta_3\exp\left[\delta_1\int \frac{ p(t)}{\varphi(t)}\,{ e^{\int\! s(t)\,{\rm d}t}}{\rm d}t\right],\quad\beta(t)= \int\! q(t)\dot\alpha(t) {\rm d}t+\delta_4.
 \end{gather*}
Here $\delta_j$ $(j=1,2,3,4)$ are arbitrary constants with $(\delta_1{}^2+\delta_2{}^2)\delta_3\not=0$.

\end{theorem}
\begin{remark}
If $r\neq0,\pm1$ is a constant, then the coefficient   $\alpha$ can be simplified as follows
$\alpha(t)=\delta_3\varphi(t)^\frac1{r-1}.$
\end{remark}

\begin{theorem}
The class of equations of the form
\begin{equation}\label{vc_KdVtx_flin}
\mathcal L_4:\quad u_t+(p(t)x+q(t))uu_x+g(t,x)u_{xxx}=0
\end{equation}
admits the generalized extended equivalence group~$\hat G^{\sim}_4$ consisting of the transformations:
\begin{gather*}
\tilde t=\theta(t),\quad \tilde x=\alpha(t)x+\beta(t),\quad\tilde u=\delta_1(u+\delta_2),  \\
\tilde p(\tilde t)=\dfrac{p(t)}{\delta_1 \dot\theta(t)} , \quad \tilde q(\tilde t)=\dfrac{1}{\delta_1\dot\theta(t)}(\alpha(t) q(t)-\beta(t) p(t)),\quad
\tilde g(\tilde t,\tilde x)=\dfrac{\alpha(t)^3}{\dot\theta(t)} g(t,x),
\end{gather*}
where~$\theta$ is an
 arbitrary  function of~$t$ with $\dot\theta\neq0$,
 \[\alpha(t)= \delta_3e^{\,\delta_2\int\!p(t)\,{\rm d}t},\quad \beta(t)={\delta_2}{\delta_3}\int\! q(t)e^{\,\delta_2\int\!p(t)\,{\rm d}t}\,{\rm d}t+\delta_4. \]
Here $\delta_j$ $(j=1,2,3,4)$ are arbitrary constants, $\delta_1\delta_3\not=0$.
\end{theorem}

\begin{theorem}
The generalized equivalence group $\hat G^\sim_5$ of the class of equations
\begin{equation}\label{vc_KdVtx_f1x}
\mathcal L_5:\quad u_t+\frac{k(t)}{x+c}\,uu_x+g(t,x)u_{xxx}=0
\end{equation}
comprises the transformations:
\begin{gather*}
\tilde t=\theta(t),\quad \tilde x=\alpha(t)(x+c)+\delta_2,\quad\tilde u=\frac{\delta_1}{\alpha(t)^2}\,u-\frac12\delta_1\delta_3 (x+c)^2,  \\
\tilde k(\tilde t)=\frac{\alpha(t)^4}{\delta_1 \dot\theta(t)}k(t), \quad \tilde c=-\delta_2,\quad
\tilde g(\tilde t,\tilde x)=\dfrac{\alpha(t)^3}{\dot\theta(t)} g(t,x),
\end{gather*}
where~$\theta$ is an
 arbitrary  function of~$t$ with $\dot\theta\neq0$,
 \[\alpha(t)= \pm\left(\delta_3\int\!k(t)\,{\rm d}t+\delta_4\right)^{-\frac12}, \]
and  $\delta_j$ $(j=1,2,3,4)$ are arbitrary constants with $\delta_1(\delta_3{}^2+\delta_4{}^2)\not=0$.
\end{theorem}

\begin{theorem}
Classes $\mathcal L_1$, $\mathcal L_2$, $\mathcal L_3$, $\mathcal L_4$, and $\mathcal L_5$ are normalized in the generalized extended sense, their equivalence groupoids are induced by transformations from the corresponding generalized extended equivalence groups $\hat G^\sim_i,$ $i=1,\dots,5$.
\end{theorem}
\begin{theorem}
The subclass $\mathcal L_0=\mathcal L\setminus \bigcup_{i=1}^5\mathcal L_i$ that is the complement of the union of the subclasses $\mathcal L_i$, $i=1,\dots, 5,$ in the class $\mathcal L$
 is normalized in the usual sense. Its equivalence group coincides with the   usual equivalence group $G^\sim$ of the whole class~$\mathcal L$.
\end{theorem}
There are no point transformations between any two equations from different subclasses~$\mathcal{L}_i$,
$i=0,\dots,5,$ of the class~\eqref{vc_KdVtx}.


\section{Equivalence groups of the gauged subclasses  $\boldsymbol{\mathcal L_1,\dots,\mathcal L_5}$}
Using equivalence transformations we can perform the gauging of the arbitrary elements depending on $t$ in each subclass ${\mathcal L_1,\dots,\mathcal L_5}$ of the class~$\mathcal L$ and therefore, to reduce the number of their arbitrary elements.
We consider each subclass separately.

$\boldsymbol{\mathcal L_1}$. The equivalence transformations with $\tilde t=\int\!f(t){\rm d}t,$ $\tilde x=x$, and $\tilde u=u$, which belong the group $\hat G^\sim_1$, reduces each equation  from class~\eqref{vc_KdVtx_fx=0} to the equation from the same class with $\tilde f=1$. Therefore, without loss of generality the gauged subclass
\begin{equation*}
\check{\mathcal L}_1:\quad u_t+uu_x+g(t,x)u_{xxx}=0
\end{equation*}
can be investigated instead of $\mathcal L_1.$ As $\mathcal L_1$ is normalized it is easy to deduce the equivalence group of its subclass $\check{\mathcal L}_1$. We simply put $\tilde f=1$ and $f=1$ in transformations from the group $\hat G^\sim_1$ and get the equation on $\theta$, $\dot\theta=\delta_1/(\delta_3t+\delta_4)^2.$ In order to write the obtained transformations in the unified form, which include both cases $\delta_3=0$ and $\delta_3\neq0$, we redenote the constants involved in transformations and get the following assertion (the corollary of Theorem~2).

\begin{corollary}The generalized extended equivalence group of the class $\check{\mathcal L}_1$ is trivial. It coincides with the usual equivalence group of this class, which is comprised of transformations
  \begin{gather*}
  \tilde t=\frac{\alpha t+\beta}{\gamma t+\delta},\quad \tilde x=\frac{\kappa x+\mu_1t+\mu_0}{\gamma t+\delta},\quad \tilde u=\frac{\kappa(\gamma t+\delta)-\kappa\gamma x+\mu_1\delta-\mu_0\gamma}{\alpha\delta-\beta\gamma},\\
 \tilde g(\tilde t,\tilde x)=\frac{\kappa^3}{\alpha\delta-\beta\gamma}\frac{g(t,x)}{\gamma t+\delta},
\end{gather*}
 $\alpha, \beta, \gamma, \delta, \kappa, \mu_1, \mu_0$ are constants defined up to a nonzero multiplier,
 $\kappa(\alpha\delta-\beta\gamma)\neq0$,  and without loss of generality we can assume that $\alpha\delta-\beta\gamma=\pm1$.

\end{corollary}
This group is in fact coincides with the equivalence group of subclass of class  $\check{\mathcal L}_1$, singled out by the condition $g_x=0$, which was derived in~\cite{popo2010b}.

$\boldsymbol{\mathcal L_2}$\,--\,$\boldsymbol{\mathcal L_4}$. Each equation from the classes ${\mathcal L_2},$ ${\mathcal L_3},$ and ${\mathcal L_4},$ is reducible to an equation from the respective class with $\tilde p=1$ by the equivalence transformation with $\tilde t=\int\!p(t)\,{\rm d}t,$ $\tilde x=x$, and $\tilde u=u$. The equivalence groups of the gauged  classes directly follow from Theorems 3--5 if we put $\tilde p=1$ and $p=1$ in the transformations therein. The respective Corollaries 2--4 are adduced in the following assertions.
\begin{corollary}
The generalized extended equivalence group of the class
\begin{equation*}\label{L2_gauged}
\check{\mathcal L}_2:\quad u_t+ e^{\,q(t)x}uu_x+\frac{s(t)- \dot q(t) x}{q(t)^3}\,u_{xxx}=0
\end{equation*}
 consists of the transformations
\begin{gather*}
\tilde t=\delta_1\int\frac1{\varphi(t)}e^{-\frac{1}{\delta_1}\beta(t)q(t)}{\rm d}t+\delta_0,\quad \tilde x=\delta_1x+\beta(t),\quad\tilde u=\varphi(t) u+\delta_3e^{-q(t)x+\int s(t)\,{\rm d}t},  \\
 \tilde q(\tilde t)=\frac {q(t)}{\delta_1},\quad \tilde s(\tilde t)=\frac1{{\delta_1}^2}{\varphi(t) e^{\frac{1}{\delta_1}\beta(t)q(t)}}\left(\delta_1s(t)+{\dot q(t)}\beta(t)\right),
\end{gather*}
where $\delta_j$ $(j=0,1,2,3,4)$ are arbitrary constants, $\delta_1(\delta_3{}^2+\delta_4{}^2)\not=0$,
 \[\beta(t)=\delta_1\delta_3\int\frac{e^{\int\! s(t)\,{\rm d}t}}{\varphi(t)}\, {\rm d}t+\delta_2,\quad \varphi(t)=\delta_3\int\! q(t)e^{\int\! s(t)\,{\rm d}t} {\rm d}t+\delta_4.\]
\end{corollary}
\begin{corollary}
The generalized extended equivalence group  of the class
\begin{gather*}
\check{\mathcal L_3}\colon\quad u_t+(x+q(t))^{r(t)}uu_x+{}\\\qquad\qquad \dfrac{(x+q(t))^{3}}{r(t)(r(t)^2-1)}\left(s(t)-\dot r(t)\ln(x+q(t))+\dfrac{(1-r(t))\dot q(t)}{x+q(t)}\right)u_{xxx}=0,
\end{gather*}
consists of the transformations
\begin{gather*}
\tilde t=\theta(t),\quad \tilde x=\alpha(t) x+\beta(t),\quad\tilde u=\varphi(t) u+\delta_1e^{\int\! s(t)\, {\rm d}t} (x+q(t))^{1-r(t)},  \\[1ex]
\tilde q(\tilde t)= q(t)\alpha(t)-  {\beta(t)},\quad
\tilde s(\tilde t)=\alpha(t)^{r(t)-1}\varphi(t)\left(s(t)+\dot r(t)\ln\alpha(t)\right),\quad\tilde r(\tilde t)=r(t),
\end{gather*}
where~$\theta$ is an
 arbitrary nonvanishing function of~$t$, $\dot\theta \not=0$,
 \begin{gather*}
 \varphi(t)=\delta_1\int\!(r(t)-1)\,e^{\int s(t){\rm d}t}{\rm d}t+\delta_2,\\
 \alpha(t)=\delta_3\exp\left[\delta_1\int \frac{{ e^{\int s(t){\rm d}t}}}{\varphi(t)}\,{\rm d}t\right],\quad\beta(t)= \int\! q(t)\dot\alpha(t) {\rm d}t+\delta_4.
 \end{gather*}
Here $\delta_j$ $(j=1,2,3,4)$ are arbitrary constants with $(\delta_1{}^2+\delta_2{}^2)\delta_3\not=0$.

\end{corollary}

\begin{corollary}
The generalized extended equivalence group $\check G^\sim_4$ of the class
\begin{equation}\label{class-l4}
\check{\mathcal L}_4:\quad u_t+(x+q(t))uu_x+g(t,x)u_{xxx}=0.
\end{equation}
 is formed by of the transformations
\begin{gather*}
\tilde t=\frac1{\delta_1}t+\delta_0,\quad \tilde x=\delta_3e^{\,\delta_2 t}x+{\delta_2}{\delta_3}\int\! q(t)e^{\,\delta_2 t}\,{\rm d}t+\delta_4,\quad\tilde u=\delta_1(u+\delta_2),  \\
\tilde q(\tilde t)=\delta_3e^{\,\delta_2 t} q(t)-{\delta_2}{\delta_3}\int\! q(t)e^{\,\delta_2 t}\,{\rm d}t-\delta_4,\quad
\tilde g(\tilde t,\tilde x)=\delta_1{\delta_3}^3e^{\, 3\delta_2 t} g(t,x),
\end{gather*}
where~$\delta_j$ $(j=0,1,2,3,4)$ are arbitrary constants, $\delta_1\delta_3\not=0$.
\end{corollary}

$\boldsymbol{\mathcal L_5}$. In the case of subclass $\mathcal L_5$ of class~\eqref{vc_KdVtx} two arbitrary elements can be gauged, namely the function $k(t)$
can be set to one, and the constant $c$ to zero. This gauge is realized by the equivalence transformation with
$\tilde t=\int\!k(t)\,{\rm d}t,$ $\tilde x=x+c$, and $\tilde u=u$. The equivalence group of the gauged class
\begin{equation*}
\check{\mathcal L}_5:\quad u_t+\frac1xuu_x+g(t,x)u_{xxx}=0
\end{equation*}
is presented in the following statement.
\begin{corollary}
The generalized extended equivalence group  of the class $\check{\mathcal L}_5$ is trivial. It coincides with the usual equivalence group of the class
comprised of the transformations
\begin{gather*}
\tilde t=\frac{\alpha t+\beta}{\gamma t+\delta},\quad \tilde x=(\gamma t+\delta)^{-\frac12}x,\quad\tilde u=\frac1{\alpha\delta-\beta\gamma}\left((\gamma t+\delta)u-\frac12\gamma x^2\right),  \\
 \tilde g(\tilde t,\tilde x)=\dfrac{(\gamma t+\delta)^\frac12}{\alpha\delta-\beta\gamma} g(t,x),
\end{gather*}
where~$\alpha,$  $\beta,$ $\gamma,$  and $\delta$   are constants defined up to a nonzero multiplier with
 $\alpha\delta-\beta\gamma\neq0$, and without loss of generality we can assume that $\alpha\delta-\beta\gamma=\pm1$.
\end{corollary}
All the gauged subclasses remain normalized, their equivalence groupoids are generated by the respective equivalence groups presented in Corollaries 1--5.
The following statement is true.
\begin{theorem}
The classes $\check{\mathcal L_1}$ and $\check{\mathcal L_5}$ are normalized in the usual sense.
The classes
 $\check{\mathcal L_2}$, $\check{\mathcal L_3}$, and $\check{\mathcal L_4}$ are normalized in the generalized extended sense.
\end{theorem}

This theorem implies that the transformational properties of classes $\check{\mathcal L_1}$ and $\check{\mathcal L_5}$ are nicer than such properties of their superclasses ${\mathcal L_1}$ and ${\mathcal L_5}$. These classes are quite convenient already to investigate them with Lie symmetry or other points of view.  The classes
 $\check{\mathcal L_2}$, $\check{\mathcal L_3}$, and $\check{\mathcal L_4}$ are still normalized only in the generalized extended sense, therefore, the links between equations in each of them are quite complicated and this courses difficulties in their investigation.
In the next section we propose a way of improvement of the transformational properties of the  class~$\check{\mathcal L_4}$.

\section{Possible improvement of transformational properties\\  via mappings between classes}
Classes of differential equations that possess generalized extended
equivalence groups are more complicated for investigation than those whose equivalence groups are usual ones.
To deal with such classes the method based on mappings between classes was proposed in~\cite{vane2009}.
This method appears to be very efficient, in particular, for solving the group classification problems.
We will illustrate the method using the example of the class
 $\check{\mathcal L_4}$.
This class can be mapped to the related class of KdV-like equations
\begin{equation}\label{l4imaged}
u_{t}+  x  u  u_{x}+h(t) u_{x}+ g(t,x) u_{xxx}=0
\end{equation}
by the family of point transformation
\begin{equation}\label{l4tr}
\hat t=t,\quad \hat x=x+q(t),\quad \hat u=u,
\end{equation}
 parameterized by the arbitrary element $q$ of the class.
The  element $h$ in the imaged  class is connected with the arbitrary element $q$ in the initial class via the formula $h(t)=\dot q(t).$
Therefore, the case $q={\rm const}$ corresponds to the case $h=0.$ In contrast to the  class $\check{\mathcal L_4}$ that is normalized in the generalized extended sense, class~\eqref{l4imaged} is normalized in the usual sense.
The equivalence groupoid of class~\eqref{l4imaged} is induced by  the transformations from its usual equivalence group $G^\sim_4$:
\begin{gather*}
\tilde t=\delta_1 t+\delta_2,\quad\tilde x=\delta_3e^{\delta_4t}x,\quad \tilde u=\frac1{\delta_1}(u+\delta_4), \\ \tilde h(\tilde t)=\frac{\delta_3}{\delta_1}e^{\delta_4t}h(t),\quad \tilde g(\tilde t,\tilde x)=\frac{\delta_3^{\,\,3}}{\delta_1}{e^{3\delta_4t}}g(t,x),
\end{gather*}
where~$\delta_j$ $(j=1,2,3,4)$ are arbitrary constants, $\delta_1\delta_3\not=0$.

The group classification of class~\eqref{l4tr} can be performed using the standard method~\cite{ovsiannikov} based on integration of the determining equations up to the $G^\sim_4$-equivalence.
The results of the group classification of equations~\eqref{l4tr} with $h\neq0$ (resp. $h=0$) are presented in Table~1 (resp. Table~2).

\begin{center}\small\renewcommand{\arraystretch}{1.4}
\setcounter{tbn}{0}
\refstepcounter{table}\label{TableLieSym1}
\textbf{Table~\thetable.}
The group classification of the class $u_{t}+  x  u  u_{x}+h(t) u_{x}+ g(t,x) u_{xxx}=0$, $hg\neq0$.
\\[2ex]
\begin{tabular}{|c|c|c|l|}
\hline
no.&Equation&Constraints&\hfil Basis of $A^{\max}$ \\
\hline
\refstepcounter{tbn}\label{TableLieSym_1.1}\thetbn&$u_t+xuu_x+ t^{m-1} u_x+t^{3m-1}\Phi (x/t^m)u_{xxx}=0$&$\Phi (z)\neq \lambda z^2$ if $m=1$&$t\partial_t+mx\partial_x-u\partial_u$\\
\refstepcounter{tbn}\label{TableLieSym_1.2}\thetbn&$u_t+xuu_x+  e^{\frac{\varepsilon}2 t^2} u_x+e^{\frac{3\varepsilon}2 t^2}\Phi \left(xe^{-\frac{\varepsilon}2 t^2}\right)u_{xxx}=0$&&$\partial_t+\varepsilon tx\partial_x+\varepsilon\partial_u$\\
\refstepcounter{tbn}\label{TableLieSym_1.3}\thetbn&$u_t+xuu_x+  u_x+\Phi (x)u_{xxx}=0$&$\Phi (x)\neq \lambda x^2$&$\partial_t$\\
\refstepcounter{tbn}\label{TableLieSym_1.4}\thetbn&$u_t+xuu_x+  u_x+\lambda x^2u_{xxx}=0$&&$\partial_t,\ t\partial_t+x\partial_x-u\partial_u$\\
\hline
\end{tabular}
\\[2ex]
\parbox{150mm}{Here $\varepsilon=\pm1\bmod\, G^\sim_4,$ $\lambda$ and $m$ are arbitrary  constants, $\lambda\neq0$.}
\end{center}

\begin{center}\small\renewcommand{\arraystretch}{1.4}
\setcounter{tbn}{0}
\refstepcounter{table}\label{TableLieSym2}
\textbf{Table~\thetable.}
The group classification of the class $u_{t}+  x  u  u_{x}+ g(t,x) u_{xxx}=0$, $g\neq0$.
\\[2ex]
\begin{tabular}{|c|c|c|l|}
\hline
no.&Equation&Constraints&\hfil Basis of $A^{\max}$ \\
\hline
\refstepcounter{tbn}\label{TableLieSym_1.5}\thetbn&$u_t+xuu_x+ \Phi(x)u_{xxx}=0$&$\Phi(x)\neq \lambda x^m$&$\partial_t$\\
\refstepcounter{tbn}\label{TableLieSym_1.6}\thetbn&$u_t+xuu_x+t^{3m-1}\Phi(x/t^m)u_{xxx}=0$&$\Phi(z)\neq \lambda z^3,\lambda z^\frac{3m-1}m$&$t\partial_t+mx\partial_x-u\partial_u$\\
\refstepcounter{tbn}\label{TableLieSym_1.7}\thetbn&$u_t+xuu_x +e^{\frac{3\varepsilon}2 t^2}\Phi\left(xe^{-\frac{\varepsilon}2 t^2}\right)u_{xxx}=0$&$\Phi(z)\neq \lambda z^3$&$\partial_t+\varepsilon tx\partial_x+\varepsilon\partial_u$\\
\refstepcounter{tbn}\label{TableLieSym_1.8}\thetbn&$u_t+xuu_x +x^m u_{xxx}=0$&$m\neq3$&$\partial_t,\   t\partial_t+\frac{1}{3-m}x\partial_x-u\partial_u$\\
\refstepcounter{tbn}\label{TableLieSym_1.9}\thetbn&$u_t+xuu_x +\Phi(t)x^3u_{xxx}=0$&$\Phi(t)\neq{\rm \lambda},\dfrac1{\lambda t+\nu}$&$x\partial_x,\ tx\partial_x+\partial_u$\\
\refstepcounter{tbn}\label{TableLieSym_1.10}\thetbn&$u_t+xuu_x +x^3u_{xxx}=0$&&$\partial_t,\ x\partial_x,\ tx\partial_x+\partial_u$\\
\refstepcounter{tbn}\label{TableLieSym_1.11}\thetbn&$u_t+xuu_x+\lambda\dfrac{x^3}tu_{xxx}=0$&&$x\partial_x, \ t\partial_t-u\partial_u,\ tx\partial_x+\partial_u$\\
\hline
\end{tabular}
\\[2ex]
\parbox{150mm}{Here $\varepsilon=\pm1\bmod\, G^\sim_4,$ $m$, $\lambda$ and $\nu$ are arbitrary constants, $\lambda\neq0.$}
\end{center}

The group classification of equations~\eqref{class-l4} with $q\ne{\rm const}$ can be recovered from the classification results obtained for equations~\eqref{l4imaged} with $h\neq0$ by using the transformation~\eqref{l4tr}. The results are presented in Table~3.
\begin{center}\small\renewcommand{\arraystretch}{1.4}
\setcounter{tbn}{0}
\refstepcounter{table}\label{TableLieSym3}
\textbf{Table~\thetable.}
The group classification of the class $u_{t}+  (x+q(t))  u  u_{x}+ g(t,x) u_{xxx}=0$, $\dot q g\neq0$.
\\[2ex]
\begin{tabular}{|c|c|c|l|}
\hline
no.&Equation&Constraints&\hfil Basis of $A^{\max}$ \\
\hline
\refstepcounter{tbn}\label{TableLieSym_3.1}\thetbn&$u_t+\left(x+t^m\right)uu_x+t^{3m-1}\Phi\left(\frac{x}{t^m}\right)u_{xxx}=0$&$\Phi(z)\neq \lambda (z+1)^2$&$t\partial_t+mx\partial_x-u\partial_u$\\
&& if $m=1$&\\
\refstepcounter{tbn}\label{TableLieSym_3.1a}\thetbn&$u_t+\left(x+\ln t\right)uu_x+t^{-1}\Phi\left({x+\ln t}\right)u_{xxx}=0$&&$t\partial_t-\partial_x-u\partial_u$\\
\refstepcounter{tbn}\label{TableLieSym_3.2}\thetbn&$u_t+\omega uu_x+e^{\frac{3\varepsilon}2 t^2}\Phi\left(\omega e^{-\frac{\varepsilon}2 t^2}\right)u_{xxx}=0$&&$\partial_t+\left(\varepsilon t\omega-e^{\frac{\varepsilon}2t^2}\right)\partial_x+\varepsilon\partial_u$\\
\refstepcounter{tbn}\label{TableLieSym_3.3}\thetbn&$u_t+(x+t)uu_x+ \Phi(x+t)u_{xxx}=0$&$\Phi(z)\neq \lambda z^2$&$\partial_t-\partial_x$\\
\refstepcounter{tbn}\label{TableLieSym_3.4}\thetbn&$u_t+(x+t)uu_x+ \lambda(x+t)^2u_{xxx}=0$&&$\partial_t-\partial_x,\ t\partial_t+x\partial_x-u\partial_u$\\
\hline
\end{tabular}
\\[2ex]
\parbox{150mm}{Here $\omega=x+\int e^{\frac{\varepsilon}2t^2} {\rm d}t$, $\varepsilon=\pm1\bmod\, \check G^\sim_4,$ $\lambda$ and $m$ are arbitrary nonzero constants. }
\end{center}

The group classification list for equations~\eqref{class-l4} with $q={\rm const}$ coincides with such a list for equations~\eqref{l4imaged} with $h=0$ and, therefore, it is presented by the list given in Table~2.

Note that in~\cite{Gungor&Lahno&Zhdanov2004} the group classification for more general class of KdV-like equations that includes
 classes~\eqref{class-l4},~\eqref{l4imaged} was carried out.  However those results obtained up to very wide equivalence group
  seem to be inconvenient to derive group classifications for classes~\eqref{class-l4} and~\eqref{l4imaged}.

\section{Conclusion}
We have presented an exhaustive study of admissible point transformations in the class of variable coefficient Korteweg--de Vries equations~\eqref{vc_KdVtx},
it has been proved that all such transformations are fiber-preserving ones.

As any contact admissible transformation between two equations from the class~\eqref{vc_KdVtx}
is the first prolongation of a~point transformation between these equations~\cite{vane2014phys},
the knowledge of the point equivalence groupoid provides us with the exhaustive description of the contact equivalence groupoid of the class.
Therefore, admissible contact transformations in the class~\eqref{vc_KdVtx} need no separate study.

We have shown that the class~\eqref{vc_KdVtx} is not normalized. Its equivalence groupoid has a complicated structure, namely, the class~\eqref{vc_KdVtx}
can be partitioned into the six disjoint normalized subclasses:

\begin{itemize}
\item the subclass $\mathcal L_0=\mathcal L\setminus \bigcup_{i=1}^5\mathcal L_i$
is normalized in the usual sense, the set of its admissible transformations is generated by point transformations from the   usual equivalence group $G^\sim$ of the entire class~\eqref{vc_KdVtx} adduced in Theorem~1;
\item the subclasses $\mathcal L_i$, $i=1,\dots,5$, which are normalized in the generalized extended sense, the sets of their admissible transformations are generated by transformations from the corresponding generalized extended groups $\hat G^\sim_i$ presented in Theorems 2--6.
\end{itemize}

\subsection*{Acknowledgments}{
SP acknowledges the support of SGS15/215/OHK4/3T/14, project of the grant agency of the Czech Technical University in Prague.
OV  would like to thank  the Department of mathematics, FNSPE, Czech Technical University in Prague for the hospitality and support.
The authors are grateful to Roman Popovych for valuable discussions, and to an  anonymous reviewer for useful remarks.}

\bibliographystyle{abbrv}
\bibliography{vcKdV}

\begin{thebibliography}{10}

\bibitem{bihlo2012b}
A.~Bihlo, E.~Dos Santos Cardoso-Bihlo, and R.~O. Popovych.
\newblock Complete group classification of a class of nonlinear wave equations.
\newblock {\em J. Math. Phys.}, 53(12):123515, 2012.

\bibitem{bihlo2015}
A.~Bihlo, E.~Dos Santos Cardoso-Bihlo, and R.~O. Popovych.
\newblock Algebraic method for finding equivalence groups.
\newblock {\em J. Phys.: Conf. Ser.}, 621(1):012001, 2015.

\bibitem{boyko2015}
V.~M. Boyko, R.~O. Popovych, and N.~M. Shapoval.
\newblock Equivalence groupoids of classes of linear ordinary differential
  equations and their group classification.
\newblock {\em J. Phys.: Conf. Ser.}, 621(1):012002, 2015.

\bibitem{brugarino}
T.~Brugarino.
\newblock Painlev\'e property, auto-{B}\"acklund transformation, {L}ax pairs,
  and reduction to the standard form for the {K}orteweg--de {V}ries equation
  with nonuniformities.
\newblock {\em J. Math. Phys.}, 30(5):1013--1015, 1989.

\bibitem{cascaval}
R.~C. Cascaval.
\newblock Variable coefficient {K}d{V} equations and waves in elastic tubes.
\newblock In {\em Evolution equations}, volume 234 of {\em Lecture Notes in
  Pure and Appl. Math.}, pages 57--69. Dekker, New York, 2003.

\bibitem{Gazeau1992}
J.-P. Gazeau and P.~Winternitz.
\newblock Symmetries of variable coefficient {K}orteweg--de {V}ries equations.
\newblock {\em J. Math. Phys.}, 33(12):4087--4102, 1992.

\bibitem{grimshaw}
R.~Grimshaw.
\newblock Slowly varying solitary waves. {I}. {K}orteweg--de {V}ries equation.
\newblock {\em Proc. Roy. Soc. London Ser. A}, 368(1734):359--375, 1979.

\bibitem{Gungor&Lahno&Zhdanov2004}
F.~G{\"u}ng{\"o}r, V.~I. Lahno, and R.~Z. Zhdanov.
\newblock Symmetry classification of {K}d{V}-type nonlinear evolution
  equations.
\newblock {\em J. Math. Phys.}, 45(6):2280--2313, 2004.

\bibitem{gungor1996}
F.~G\"ung\"or, M.~Sanielevici, and P.~Winternitz.
\newblock On the integrability properties of variable coefficient
  {K}orteweg--de {V}ries equations.
\newblock {\em Can. J. Phys.}, 74(9-10):676--684, 1996.

\bibitem{gurses}
M.~G{\"u}rses and A.~Karasu.
\newblock Variable coefficient third order {K}orteweg-de {V}ries type of
  equations.
\newblock {\em J. Math. Phys.}, 36(7):3485--3491, 1995.

\bibitem{hlavaty}
L.~Hlavat{\'y}.
\newblock Painlev\'e analysis of nonautonomous evolution equations.
\newblock {\em Phys. Lett. A}, 128(6-7):335--338, 1988.

\bibitem{ivan2005}
N.~M. Ivanova, R.~O. Popovych, and C.~Sophocleous.
\newblock Conservation laws of variable coefficient diffusion--convection
  equations.
\newblock {\em in: N.H. Ibragimov et al. (ed.), Proc. of Tenth International
  Conference in Modern Group Analysis (Larnaca, Cyprus, 2004), Nicosia (Kyiv:
  Institute of Mathematics)}, pages 107--113, 2005.

\bibitem{jeffrey}
A.~Jeffrey and T.~Kakutani.
\newblock Weak nonlinear dispersive waves: {A} discussion centered around the
  {K}orteweg-de {V}ries equation.
\newblock {\em SIAM Rev.}, 14:582--643, 1972.

\bibitem{jeffrey_book}
A.~Jeffrey and T.~Taniuti.
\newblock {\em Non-linear wave propagation. {W}ith applications to physics and
  magnetohydrodynamics}.
\newblock Academic Press, New York-London, 1964.

\bibitem{joshi}
N.~Joshi.
\newblock Painlev\'e property of general variable-coefficient versions of the
  {K}orteweg-de {V}ries and nonlinear {S}chr\"odinger equations.
\newblock {\em Phys. Lett. A}, 125(9):456--460, 1987.

\bibitem{king1991a}
J.~G. Kingston.
\newblock On point transformations of evolution equations.
\newblock {\em J.~Phys.~A: Math. Gen.}, 24:L769--L774, 1991.

\bibitem{kingston1991}
J.~G. Kingston and C.~Sophocleous.
\newblock On point transformations of a generalised {B}urgers equation.
\newblock {\em Phys. Lett. A}, 155(1):15--19, 1991.

\bibitem{kingston1997}
J.~G. Kingston and C.~Sophocleous.
\newblock On form-preserving point transformations of partial differential
  equations.
\newblock {\em J. Phys. A: Math. Gen.}, 31(6):1597--1619, 1998.

\bibitem{kundu}
A.~Kundu.
\newblock Integrable nonautonomous nonlinear {S}chr\"odinger equations are
  equivalent to the standard autonomous equation.
\newblock {\em Phys. Rev. E (3)}, 79(1):015601, 2009.

\bibitem{kuri2015}
O.~Kuriksha, S.~Po{\v{s}}ta, and O.~Vaneeva.
\newblock Group classification of variable coefficient generalized {K}awahara
  equations.
\newblock {\em J. Phys. A}, 47(4):045201, 19, 2014.

\bibitem{kuru2015}
C.~Kurujyibwami.
\newblock Equivalence groupoid for (1+2)-dimensional linear {S}chr\"odinger
  equations with complex potentials.
\newblock {\em J. Phys.: Conf. Ser.}, 621(1):012008, 2015.

\bibitem{mele94Ay}
S.~V. Meleshko.
\newblock Group classification of equations of two-dimensional gas motions.
\newblock {\em Prikl. Mat. Mekh.}, 58(4):56--62, 1994.

\bibitem{ovsiannikov}
L.~V. Ovsiannikov.
\newblock {\em Group analysis of differential equations}.
\newblock Academic Press, Inc., New York-London, 1982.

\bibitem{poch2015}
O.~A. Pocheketa.
\newblock Equivalence groupoid of generalized potential {B}urgers equations.
\newblock {\em J. Phys.: Conf. Ser.}, 621(1):012011, 2015.

\bibitem{poch2014}
O.~A. Pocheketa, R.~O. Popovych, and O.~O. Vaneeva.
\newblock Group classification and exact solutions of variable-coefficient
  generalized {B}urgers equations with linear damping.
\newblock {\em Appl. Math. Comput.}, 243:232--244, 2014.

\bibitem{popo2006}
R.~O. Popovych.
\newblock Classification of admissible transformations of differential
  equations.
\newblock {\em {\it Collection of Works of Institute of Mathematics\/} (Kyiv:
  Institute of Mathematics)}, 3(2):239--254, 2006.

\bibitem{popo2012a}
R.~O. Popovych and A.~Bihlo.
\newblock Symmetry preserving parameterization schemes.
\newblock {\em J. Math. Phys.}, 53(7):073102, 2012.

\bibitem{popo2010a}
R.~O. Popovych, M.~Kunzinger, and H.~Eshraghi.
\newblock Admissible transformations and normalized classes of nonlinear
  {S}chr\"odinger equations.
\newblock {\em Acta Appl. Math.}, 109(2):315--359, 2010.

\bibitem{popo2010b}
R.~O. Popovych and O.~O. Vaneeva.
\newblock More common errors in finding exact solutions of nonlinear
  differential equations: {P}art {I}.
\newblock {\em Commun. Nonlinear Sci. Numer. Simul.}, 15(12):3887--3899, 2010.

\bibitem{soph_thesis}
C.~Sophocleous.
\newblock Transformation methods in the study of nonlinear partial differential
  equations. {P}h{D} {T}hesis. {U}niversity of {N}ottingham.
\newblock 1991.

\bibitem{vane2009}
O.~O. Vaneeva, R.~O. Popovych, and C.~Sophocleous.
\newblock Enhanced group analysis and exact solutions of variable coefficient
  semilinear diffusion equations with a power source.
\newblock {\em Acta Appl. Math.}, 106(1):1--46, 2009.

\bibitem{vane2014phys}
O.~O. Vaneeva, R.~O. Popovych, and C.~Sophocleous.
\newblock Equivalence transformations in the study of integrability.
\newblock {\em Physica Scripta}, 89(3):038003, 2014.

\bibitem{wint1992}
P.~Winternitz and J.-P. Gazeau.
\newblock Allowed transformations and symmetry classes of variable coefficient
  {K}orteweg--de {V}ries equations.
\newblock {\em Phys. Lett. A}, 167(3):246--250, 1992.

\end{thebibliography}

\end{document}